\documentclass[]{aa}
\usepackage{txfonts}
\usepackage{natbib}
\usepackage{amssymb}
\bibpunct{(}{)}{;}{a}{}{,}
\usepackage{epsf}
\usepackage{graphicx}
\usepackage{epsfig}
\usepackage{graphics}
\usepackage{subfigure}
\usepackage[english]{babel}
\usepackage{rotating}
\usepackage{color}

\newcommand{\teff}{$T_{\rm{eff}}$}
\newcommand{\logg}{$\log g$}
\newcommand{\lL}{\ifmmode \log \frac{L}{L_{\sun}} \else $\log \frac{L}{L_{\sun}}$\fi}

\newcommand{\vsini}{$V$~sin$i$}
\newcommand{\vturb}{$v_{\rm turb}$}

\newcommand{\kms}{km~s$^{-1}$}

\begin{document}

\title{Synthetic photometry of globular clusters:\\
  Uncertainties on synthetic colors}
\author{F. Martins\inst{1}
}
\institute{LUPM, Universit\'e de Montpellier, CNRS, Place Eug\`ene Bataillon, F-34095 Montpellier, France  \\
%           \email{fabrice.martins@umontpellier.fr}
}

\offprints{Fabrice Martins\\ \email{fabrice.martins@umontpellier.fr}}

\date{Received / Accepted }

\abstract
{Synthetic photometry is a great tool for studying globular clusters, especially for understanding the nature of their multiple populations.}
{Our goal is to quantify the errors on synthetic photometry that
are caused by uncertainties on stellar and observational/calibration parameters. These errors can be taken into account when building synthetic color-magnitude diagrams (CMDs) that are to be  compared to observed CMDs. }
{We have computed atmosphere models and synthetic spectra for two stars, Pollux and Procyon, that have stellar parameters typical of turn-off and bottom red giant branch stars in globular clusters. We then varied the effective temperature, surface gravity, microturbulence, the carbon, nitrogen, and oxygen abundances, and [$\frac{Fe}{H}$]. We quantified the effect on synthetic photometry in the following filters: Johnson $UBVRI$ and HST F275W, F336W, F410M, F438W, F555W, F606W, and F814W. We also estimated the effects of extinction, atmospheric correction, and of the Vega reference spectrum on the resulting photometry. In addition, we tested the ability of our models to reproduce the observed spectral energy distribution and observed photometry of the two stars.}
{We show that variations are generally stronger in blue filters, especially those below 4500 \AA. Dispersions on synthetic colors due to uncertainties on stellar parameters vary between less than 0.01 and to 0.04 magnitude, depending on the choice of filters. Uncertainties on the zero points, the extinction law, or the atmospheric correction affect the resulting colors at a level of a few 0.01 magnitudes in a systematic way. The models reproduce
the flux-calibrated spectral energy distribution of both stars
well. Comparison between synthetic and observed $UBVRI$ photometry shows a variable degree of (dis)agreement. The observed differences indicate that different reduction and calibration processes are
performed to obtain respectively observed and synthetic photometry, and they
call for publication of all the details of the reduction process to produce synthetic photometry at a 0.01 mag level, which is required to interpret observed CMDs.}
{}


\authorrunning{F. Martins}
\titlerunning{Uncertainties on synthetic colors of globular clusters}

\maketitle

%%%%%%%%%%%%%%%%%%%%%%%%%%%%%%%%%%%%%%%%%%%%%%%%%%%%%%%%%%%%%%%%%%%%%%%%%%%%%%%%%%%%%%%%%%%%%%%%%%%%%%%%%%%%%%%%%%%%%%%%%%%%%%%
%%%%%%%%%%%%%%%%%%%%%%%%%%%%%%%%%%%%%%%%%%%%%%%%%%%%%%%%%%%%%%%%%%%%%%%%%%%%%%%%%%%%%%%%%%%%%%%%%%%%%%%%%%%%%%%%%%%%%%%%%%%%%%%
\section{Introduction}
\label{s_intro}

Globular clusters were once known to be simple structures made of stars formed at the same time with the same initial chemical composition. This picture has been deeply revised since various sub-groups of stars have been discovered in the vast majority of them. These sub-groups, also known as multiple populations, are detected in both spectroscopy and photometry. Determinations of surface chemical abundances indicate that some stars are enriched in nitrogen, sodium, and aluminum, while being at the same time depleted in carbon, oxygen, and magnesium \citep[e.g.,][]{sneden92,kraft97,car10}. A wide range of enrichment or depletion is usually observed, leading to so-called anticorrelations between nitrogen and carbon, sodium and oxygen, and aluminum and magnesium \citep[][]{yong06,car06,gratton07,car09,marino11,car15}.
Additionally, color-magnitude diagrams (CMDs) of essentially all the globular clusters reveal multiple sequences (or at least spreads) in one or several branch (main sequence, MS; turn-off,
TO; red giant branch, RGB;  asymptotic giant branch, AGB; and
horizontal branch, HB). The Hubble Space Telescope has pioneered the identification of such sequences \citep[e.g.,][]{bedin04,piotto07,milone10,milone12,piotto15,soto17}, but they are now observed with any high spatial resolution photometric facilities \citep{han09,gruy17}.

The origin of the multiple populations observed in globular clusters remains unknown. The chemical abundance patterns all point to nucleosynthesis through the CNO cycle, Ne--Na and Mg--Al chains at high temperature \citep[75 MK,][]{pr07,pr17}. These conditions are encountered in the core of MS massive, very massive and super-massive stars or in the envelope of some AGB stars. This has led to a generation of scenarios invoking a first generation of stars formed from pristine gas. Out of this first generation, some stars \citep[massive or AGB stars;][]{ventura01,dec07,dh14,gieles18} ejected processed material that was subsequently mixed with gas to form a second generation of stars. Depending on the degree of mixing, the stars of the second generation show the observed chemical anticorrelations. The different scenarios proposed to explain the presence of multiple populations partly rely on nucleosynthesis through the CNO cycle and the Ne-Na and Mg-Al chains. As such, they also predict some degree of helium enrichment, which should be observed in stars that formed out of the ejecta of the first-generation stars. When AGB stars are the main polluters, a maximum helium mass fraction of 0.38 is expected \citep{ventura13}, while for scenarios involving massive stars, higher values are not forbidden \citep{chantereau16} and can be limited to 0.4 in the case of super-massive stars if stellar winds are efficient enough \citep{dh14}.  However, spectroscopic determinations of the helium content in globular clusters are almost impossible owing to the absence of spectroscopic features in most stars, except for hot HB objects \citep{marino14}. For the latter, complications due to atomic diffusion render abundance determinations uncertain.

Hence, determinations of the helium content of globular clusters stars have mostly been made based on an indirect method: the comparison of theoretical isochrones built with different Y (i.e., helium mass fraction) to observed CMDs. A larger helium content decreases the envelope opacity and increases the mean molecular weight, two effects that combine to make helium-rich isochrones bluer \citep[e.g.,][]{chantereau16}. The method requires the transformation of theoretical Hertzsprung-Russell diagrams into CMDs. This can be done either by direct calculations of synthetic spectra along isochrones, or by use of bolometric corrections \citep[][]{milone13,milone17}.
Most determinations of Y performed so far rely on the color differences between multiple populations: the observed differences in colors between two populations are compared to the color differences between isochrones with different Y. In that sense, these determinations provide an estimate of the \textit{\textup{relative}} helium content between multiple populations. Assuming a value of Y for the less chemically processed population (the first generation
or population), this provides an absolute value for Y for each population. Such a differential analysis usually does not take into account any dispersion in theoretical isochrones: they are plotted as single lines in CMDs. A more physical approach would be to introduce a distribution of colors around the average value of the theoretical isochrone and to take this dispersion into account when performing comparisons to observed populations in CMDs. This would not affect the determination of the Y difference when Y is significantly different between two populations. However, this may be important for small Y differences, when the overlap between two theoretical isochrones due to dispersion is non-negligible.

Another method for constraining the Y content would be to directly compare the position of theoretical isochrones to observed CMDs. This direct approach is more complex than the differential one since it involves uncertainties in the modeling of stellar evolution and atmosphere models, uncertainties that mostly cancel out in a differential approach. However, direct comparisons of isochrones to CMDs do dot require any assumption on the chemical composition of the first population.
Directly comparing theoretical isochrones to observed CMDs is also important to constrain the age of globular clusters. Again, a dispersion around theoretical isochrones must be taken into
account, however, to correctly estimate uncertainties on ages. Finally, direct comparisons are useful for testing the physics of evolutionary models and atmosphere models.

In this paper, we present an investigation of the dispersion around theoretical isochrones. Our final goal is to produce theoretical CMDs that can be directly compared to observed CMDs. We plan to produce such theoretical CMDs by drawing artificial stars with parameters centered around those of theoretical isochrones and with a distribution characterized by the uncertainties determined in this work. This should provide an independent view of the properties of globular clusters.
In Sect.\ \ref{s_method} we describe our method and the standard stars we selected. Sect.\ \ref{s_res} describes our results,
which are summarized in Sect.\ \ref{s_conc}.

%%%%%%%%%%%%%%%%%%%%%%%%%%%%%%%%%%%%%%%%%%%%%%%%%%%%%%%%%%%%%%%%%%%%%%%%%%%%%%%%%%%%%%%%%%%%%%%%%%%%%%%%%%%%%%%%%%%%%%%%%%%%%%%
\section{Method}
\label{s_method}

\begin{figure}[t]
\centering
\includegraphics[width=9cm]{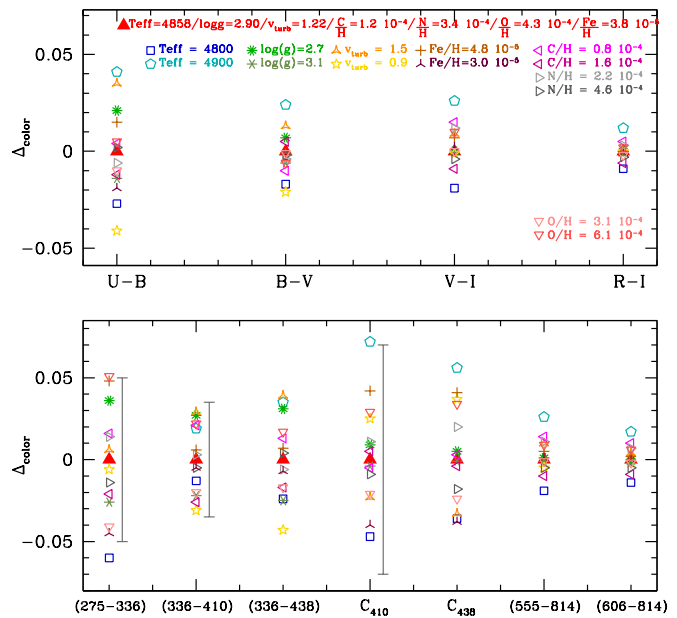}
\caption{Difference between colors of Pollux models with one parameter varied compared to the color of the reference model (red filled triangle). The parameters of the reference model are given in the upper panel, together with the values of the parameters that were varied. Colors based on Johnson photometry (HST WFC3 and ACS) are shown in the upper (lower) panel. C$_{X}$=(275-336)-(336-X), where numbers refer to magnitudes in a given filter (i.e., 275 is the magnitude in the F275W filter), and $X$ is either the F410W or the F438W filter. Gray vertical bars indicate the typical separation between RGB populations in the cluster NGC~6752, according to \citet{milone13}.}
\label{mag_pollux}
\end{figure}

\begin{figure}[t]
\centering
\includegraphics[width=9cm]{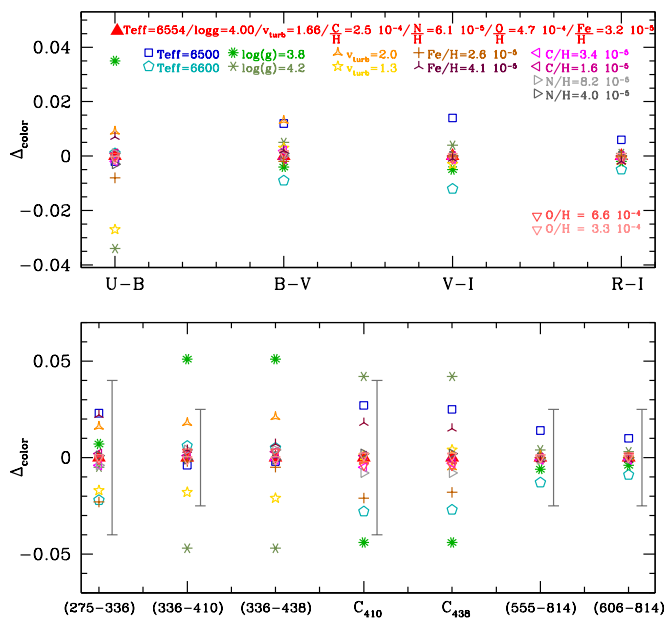}
\caption{Same as bottom panels of Fig.\ \ref{mag_pollux} for Procyon. The gray solid vertical lines in the bottom panel indicate the width of the turn-off of NGC~6752 in the corresponding color, from the data of \citet{milone13}.}
\label{mag_procyon}
\end{figure}

To estimate the dispersion around a theoretical isochrone in CMDs, we need to constrain the color variations that are due to changes in fundamental parameters and surface abundances. We assume that such variations exist in a population that is
theoretically represented by a single isochrone.

\citet{sbo11} have studied the effects of variations of various surface abundances on CMDs. They reported that C, N, and O significantly affect the shape of spectra below $\sim$ 4500 \AA. Conversely, helium has little effect on synthetic spectra at a given effective temperature, but affects the internal structure (see above) and thus \teff. As a consequence, the helium content also affects the shape of theoretical isochrones in CMDs through its effect on effective temperature. This was confirmed by \citet{cas17}, who have quantified the displacement of isochrones that is due to Y changes in synthetic CMDs built from the HST filters F606W and F814W. We thus consider C, N, O, and He as the main sources of color variations that are due to changes in surface abundances. We also take into account color variations that are due to fundamental parameters: effective temperature, surface gravity, and microturbulence. For these parameters, we assume that the dispersion in fundamental parameters is similar to the uncertainties of spectroscopic determination of such parameters. Another assumption could be to estimate the dispersion between isochrones that is produced by different groups and stellar evolution codes. We prefer using spectroscopic determinations as the source of uncertainties since they do not depend in the degree of refinement of the physics included in evolutionary models.
We also provide an estimate of some sources of systematic uncertainties on synthetic photometry: the effect of calibration, extinction, and airmass (for ground-based observations).

%------------------
\subsection{Selection of stars and stellar parameters}
\label{s_targets}

For our purpose, we first focused on RGB stars since these objects are bright and thus more easily observed in globular clusters. Spectroscopic data are available for abundance determinations. In addition, we concentrated on stars at the bottom of the RGB to avoid additional complications due to stellar evolution in more advanced phases (dredge-up and deep mixing). From these criteria, and considering only bright targets with robust photometry, we selected the K0~III star POLLUX ($\beta$~Gem, HD~62509, HR~2990) as representative of this class of objects. Its photometry is stable over time \citep{gray14}, and it is usually considered an RGB star with low luminosity. \citet{auriere15} detected a weak magnetic field of 0.5 G at its surface.

To model the spectral energy distribution, we adopted the effective temperature and surface gravity of \citet{heiter15}. We chose a value of microturbulent velocity $\xi_{t}$ of 1.22 \kms\ from \citet{luck15}. The surface abundances were taken from \citet{luck15} and \citet{jofre15a}. A projected rotational velocity (\vsini) of 2.8 \kms\ was adopted from \citet{auriere15}. \citet{gray14} provides references for the different values of the stellar parameters encountered in the literature, and we refer to this work for further information. We extract from this work the typical uncertainties: 50 to 100 K for \teff\ with modern values closer to 50 K, 0.3 dex on \logg, and 0.3 \kms\ on $\xi_{t}$. Uncertainties on surface abundances depend on the element and are listed in \citet{luck15} and \citet{jofre15a}. They are on the order of 0.10-0.15 dex in units of 12+$\log(\frac{X}{H})$. Consequently, we adopt the following errors: 0.15 dex for carbon, nitrogen, and oxygen \citep[see also][]{adam13}, and 0.10 dex for iron.

In addition to Pollux, we also considered Procyon ($\alpha$ CMi, HD1421, HR~2943), an F5V-IV star with parameters typical of TO stars in globular clusters. Multiple populations in globular clusters are less easily detected on the TO, but they probably contribute to a widening of this region of the CMD since multiple populations are observed both on the MS and in evolved phases (RGB and AGB). A good knowledge of the uncertainties affecting synthetic photometry is crucial for quantitative determinations of stellar ages.

The effective temperature and surface gravity of Procyon were adopted from \citet{heiter15}, the surface abundances from \citet{jofre15a,jofre15b}. The projected rotational velocity (2.8 \kms) and microturbulent velocity (1.66 \kms) were taken from \citet{jofre15a}. 

The adopted stellar parameters for Pollux and Procyon are given in the first line (below the star name) in Table \ref{param_colors}. The corresponding models are referred to as the ``reference models'' in the remainder of the paper.

Pollux and Procyon both have roughly solar metallicities, while stars in most globular clusters have [$\frac{Fe}{H}$] between 0.0 and -2.5 \citep{car09fe}. As stated above, Pollux and Procyon are nearby and relatively standard stars with well-determined stellar parameters and surface abundances. Finding such stars with [$\frac{Fe}{H}$]$\sim$-2.0 is difficult, since they are fainter and thus do not have spectroscopic parameters as good
as close-by objects. However, from the point of view of the determination of stellar parameters from spectroscopy, the only difference between solar metallicity and metal-poor stars is stronger non-local
thermal equilibrium (non-LTE) effects in the latter case \citep[e.g.,][]{lind12}. This adds a systematic uncertainty on stellar parameters and surface abundances, with a magnitude that increases at lower [$\frac{Fe}{H}$] \citep{merle11,ruchti13}. The statistical uncertainties (due to statistical uncertainties on \teff, \logg,\ and surface abundances) remain the same, however. Hence, this study strictly speaking applies to the most metal-rich globular clusters. At lower [$\frac{Fe}{H}$], systematic trends on colors are to be expected, in addition to the effects discussed in Sect.\ \ref{s_Vega}.

%------------------
\subsection{Atmosphere models and synthetic photometry}

We have used the atmosphere code ATLAS12 \citep{kur14} and the spectral synthesis code SYNTHE \citep{kur05} to compute the spectral energy distribution (SED). Photometry in various filters was subsequently calculated from the SED. To do this, we retrieved the Johnson $UBVRI$ filter throughputs from the General Catalogue of Photometric Data \footnote{\url{http://obswww.unige.ch/gcpd/gcpd.html}} \citep[GCPD,][]{merm97}. We also used the Spanish Virtual Observatory\footnote{\url{http://svo2.cab.inta-csic.es/svo/theory/fps3/}} to retrieve the HST/WFC3/UVIS2 filters F275W, F336W, F410M, F438W, and F555W and the HST/ACS WFC filters F606W and F814W for a temperature of -81$^{\circ}$C. For each filter, we convolved the synthetic SED with the filter throughput and calculated the corresponding flux, which was subsequently divided by the zero-point flux to give the synthetic magnitude. 

To ensure consistency in our photometry, we recalculated the zero-point fluxes for all filters in the VEGAMAG system. For this purpose, we retrieved the reference spectrum of Vega used in HST calibrations from \url{ftp://ftp.stsci.edu/cdbs/current_calspec/}. We used the spectrum ``alpha\_lyr\_stis\_008.fits'' (see also Sect.\ \ref{s_Vega}).

%%%%%%%%%%%%%%%%%%%%%%%%%%%%%%%%%%%%%%%%%%%%%%%%%%%%%%%%%%%%%%%%%%%%%%%%%%%%
%%%%%%%%%%%%%%%%%%%%%%%%%%%%%%%%%%%%%%%%%%%%%%%%%%%%%%%%%%%%%%%%%%%%%%%%%%%%
\section{Results}
\label{s_res}

%%%%%%%%%%%%%%%%%%%%%%%%%%%%%%%%%%%%%%%%%%%%%%%%%%%%%%%%%%%%%%
%
\subsection{Estimates of uncertainties on synthetic photometry}
\label{s_unc}

%------------------------------------------
\subsubsection{Effect of stellar parameters}
\label{s_effparam}
We first studied the effect of variations in stellar parameters on the resulting photometry. We  selectively varied the effective temperature, the surface gravity, the microturbulent velocity, and the abundances of carbon, nitrogen, oxygen, and iron. We focussed on carbon, nitrogen, and oxygen since they show a wide range of values in globular clusters and affect the SEDs of globular cluster stars most \citep[e.g.,][]{sbo11}. For each parameter, we selected two values bracketing the reference value listed in Sect.\ \ref{s_targets}. These new values correspond to the reference value plus/minus the uncertainty. For instance, the reference value for \teff\  for Pollux is 4858 K, with an uncertainty of about 50K. We thus ran two models with \teff\ = 4800 and 4900 K, respectively. The results are gathered in Table \ref{param_colors}.

Fig.\ \ref{mag_pollux} shows a graphical representation of some results for Pollux. In the upper panel, the dispersion in color difference is largest in the blue ($U-B$ color), where the effects of effective temperature and microturbulence are the strongest. A difference of 0.04 mag is not unexpected. Table \ref{sig_colors} gathers the dispersion in colors shown in Fig.\ \ref{mag_pollux} and \ref{mag_procyon}. The dispersion is the standard deviation of the 15 models computed for each star. For Pollux, it is 0.022 in ($U-B$). The red part of the spectrum ($R-I$ color) is less sensitive to parameter variations with color differences not larger than 0.01 magnitudes (dispersion 0.005). For the $B$ and $V$ filters, color variations are intermediate, with differences reaching 0.02 magnitude and a dispersion of 0.011 ($B-V$).

The lower panel of Fig.\ \ref{mag_pollux} shows the effects of stellar parameter variations on colors based on HST photometry for Pollux. As above, the changes are greatest in the blue part of the spectrum. For the selected filters, the color differences can reach 0.05 magnitudes. Colors involving the filters F275W, F336W, F410M, and F438W are the most affected by variations in stellar parameters. The dispersion is 0.041 for (275-336)\footnote{The notation (275-336) stands for the magnitude difference between the F275W and F336W filters. Similar notations are used for the other HST filters.} and drops to 0.008 for (606-814). These variations are important in the context of understanding multiple populations in globular clusters since photometry based on two or three of the blue filters is the most efficient in separating multiple populations \citep[][]{milone13,piotto15}. As an illustration, we show in Fig.\ \ref{mag_pollux} some color separations between multiple populations in the globular cluster NGC~6752, which is one of the best-studied clusters \citep{yong05,car05,car07,yong08,villa09,milone10,car12,charb13,car13,krav14,yong15,dotter15,nardiello15,lapenna16,muc17}. The range of parameters we explored leads to a range of colors similar to the typical color difference between populations ``a'' and ``b'' in NGC~6752 according to \citet{milone13}, see their Fig.~12. However, the dispersion formally remains below the color difference between two populations (for the case of NGC~6752 taken as reference here). For instance, the dispersion in the C$_{410}$ index is 0.031, while the difference between the two main populations is on the order 0.140 mag. In the particular case here, when a theoretical CMD is built by drawing artificial stars with parameters centered on the isochrones that best fit the two populations a and b, and when a dispersion around these theoretical isochrones is included, most artificial stars are part of two groups that are well separated in color (dispersion of 0.031 mag versus observed separation of 0.140 mag), although some of the artificial stars from the bluest population may be located at the position of the redder population (total range of colors as wide as the separation between populations a and
b).

If the separation between populations a and b were on the order of the theoretical dispersion (0.031 mag), it would have been difficult to infer the difference in properties of the two populations from the theoretical isochrones because the two artificial populations
overlap significantly; this problem does not exist when no dispersion around isochrones is considered.

\begin{sidewaystable*}
\begin{center}
\caption{Absolute magnitudes for the Pollux and Procyon models and effect of stellar parameters.}
\label{param_colors}
\begin{tabular}{lcccccccccccccccccc}
\hline
\teff  & \logg & \vturb  & C/H            & N/H            & O/H            & Fe/H     &    $U$ &  $B$   &  $V$   &  $R$    &  $I$  & F275W & F336W & F410M & F438W & F555W & F606W & F814W \\    
 $[K]$ &       & [\kms]  & [$\times 10^4$] & [$\times 10^4$] & [$\times 10^4$] & [$\times 10^5$] &   &        &        &        &   & & & & & & &    \\
\hline
 & & & & &  & & & Pollux & \\
\hline
4858   & 2.90  &  1.22 &  1.2             & 3.4       & 4.3      & 3.8     &  2.308  & 1.817  & 0.910  & 0.319  & -0.139  &  4.340  &  2.314  &  2.229  &  1.934  &  1.063  &  0.708  & -0.034 \\
4800   & 2.90  &  1.22 &  1.2             & 3.4       & 4.3      & 3.8     &  2.451  & 1.919  & 0.988  & 0.383  & -0.087  &  4.573  &  2.456  &  2.352  &  2.041  &  1.144  &  0.780  &  0.021 \\
4900   & 2.90  &  1.22 &  1.2             & 3.4       & 4.3      & 3.8     &  2.209  & 1.745  & 0.855  & 0.274  & -0.175  &  4.181  &  2.215  &  2.143  &  1.859  &  1.006  &  0.656  & -0.072 \\
4858   & 2.70  &  1.22 &  1.2             & 3.4       & 4.3      & 3.8     &  2.335  & 1.823  & 0.909  & 0.318  & -0.140  &  4.415  &  2.353  &  2.241  &  1.942  &  1.063  &  0.706  & -0.035 \\
4858   & 3.10  &  1.22 &  1.2             & 3.4       & 4.3      & 3.8     &  2.289  & 1.812  & 0.911  & 0.320  & -0.138  &  4.283  &  2.283  &  2.220  &  1.928  &  1.063  &  0.709  & -0.033 \\
4858   & 2.90  &  1.50 &  1.2             & 3.4       & 4.3      & 3.8     &  2.363  & 1.837  & 0.917  & 0.316  & -0.141  &  4.407  &  2.375  &  2.261  &  1.956  &  1.071  &  0.710  & -0.036 \\
4858   & 2.90  &  0.90 &  1.2             & 3.4       & 4.3      & 3.8     &  2.248  & 1.798  & 0.912  & 0.322  & -0.136  &  4.267  &  2.247  &  2.193  &  1.910  &  1.062  &  0.710  & -0.030 \\
4858   & 2.90  &  1.22 &  1.6             & 3.4       & 4.3      & 3.8     &  2.292  & 1.813  & 0.901  & 0.313  & -0.139  &  4.299  &  2.294  &  2.235  &  1.931  &  1.053  &  0.699  & -0.034 \\
4858   & 2.90  &  1.22 &  0.8             & 3.4       & 4.3      & 3.8     &  2.318  & 1.823  & 0.926  & 0.325  & -0.138  &  4.370  &  2.328  &  2.222  &  1.935  &  1.078  &  0.719  & -0.033 \\
4858   & 2.90  &  1.22 &  1.2             & 4.6       & 4.3      & 3.8     &  2.305  & 1.812  & 0.907  & 0.317  & -0.137  &  4.324  &  2.312  &  2.232  &  1.928  &  1.059  &  0.704  & -0.033 \\
4858   & 2.90  &  1.22 &  1.2             & 2.2       & 4.3      & 3.8     &  2.310  & 1.825  & 0.922  & 0.322  & -0.139  &  4.354  &  2.314  &  2.226  &  1.940  &  1.074  &  0.715  & -0.034 \\
4858   & 2.90  &  1.22 &  1.2             & 3.4       & 6.1      & 3.8     &  2.317  & 1.821  & 0.920  & 0.321  & -0.139  &  4.409  &  2.332  &  2.225  &  1.935  &  1.072  &  0.714  & -0.034 \\
4858   & 2.90  &  1.22 &  1.2             & 3.4       & 3.1      & 3.8     &  2.298  & 1.817  & 0.911  & 0.318  & -0.138  &  4.281  &  2.296  &  2.231  &  1.933  &  1.063  &  0.707  & -0.033 \\
4858   & 2.90  &  1.22 &  1.2             & 3.4       & 4.3      & 4.8     &  2.327  & 1.821  & 0.915  & 0.317  & -0.141  &  4.397  &  2.323  &  2.232  &  1.936  &  1.066  &  0.708  & -0.036 \\
4858   & 2.90  &  1.22 &  1.2             & 3.4       & 4.3      & 3.0     &  2.290  & 1.818  & 0.915  & 0.321  & -0.136  &  4.287  &  2.306  &  2.226  &  1.933  &  1.067  &  0.711  & -0.031 \\
\hline
 & & & & &  & & & Procyon & \\
\hline
6554 & 4.00 & 1.66 &  2.5             & 0.6     & 4.7     & 3.2   &  2.935 &  3.017 &  2.605 &  2.302 &  2.079  & 3.795 & 2.897 & 3.162 & 3.060 & 2.686 & 2.507 & 2.121 \\
6600 & 4.00 & 1.66 &  2.5             & 0.6     & 4.7     & 3.2   &  2.894 &  2.975 &  2.572 &  2.276 &  2.058  & 3.734 & 2.858 & 3.117 & 3.016 & 2.651 & 2.476 & 2.099 \\
6500 & 4.00 & 1.66 &  2.5             & 0.6     & 4.7     & 3.2   &  2.984 &  3.068 &  2.644 &  2.333 &  2.104  & 3.867 & 2.946 & 3.215 & 3.111 & 2.726 & 2.543 & 2.147 \\
6554 & 3.80 & 1.66 &  2.5             & 0.6     & 4.7     & 3.2   &  2.960 &  3.007 &  2.599 &  2.299 &  2.078  & 3.842 & 2.937 & 3.151 & 3.049 & 2.679 & 2.502 & 2.120 \\
6554 & 4.20 & 1.66 &  2.5             & 0.6     & 4.7     & 3.2   &  2.911 &  3.027 &  2.610 &  2.304 &  2.080  & 3.753 & 2.860 & 3.172 & 3.070 & 2.691 & 2.511 & 2.122 \\
6554 & 4.00 & 2.50 &  2.5             & 0.6     & 4.7     & 3.2   &  2.983 &  3.023 &  2.599 &  2.292 &  2.071  & 3.893 & 2.954 & 3.176 & 3.067 & 2.680 & 2.499 & 2.113 \\
6554 & 4.00 & 0.00 &  2.5             & 0.6     & 4.7     & 3.2   &  2.966 &  3.045 &  2.625 &  2.318 &  2.092  & 3.844 & 2.930 & 3.192 & 3.088 & 2.707 & 2.526 & 2.134 \\
6554 & 4.00 & 1.66 &  3.4             & 0.6     & 4.7     & 3.2   &  2.933 &  3.017 &  2.603 &  2.300 &  2.078  & 3.790 & 2.896 & 3.160 & 3.060 & 2.684 & 2.505 & 2.120 \\
6554 & 4.00 & 1.66 &  1.6             & 0.6     & 4.7     & 3.2   &  2.936 &  3.017 &  2.606 &  2.303 &  2.080  & 3.799 & 2.899 & 3.163 & 3.059 & 2.687 & 2.508 & 2.122 \\
6554 & 4.00 & 1.66 &  2.5             & 0.8     & 4.7     & 3.2   &  2.936 &  3.017 &  2.605 &  2.301 &  2.079  & 3.794 & 2.900 & 3.161 & 3.059 & 2.685 & 2.507 & 2.121 \\
6554 & 4.00 & 1.66 &  2.5             & 0.4     & 4.7     & 3.2   &  2.933 &  3.018 &  2.605 &  2.302 &  2.079  & 3.795 & 2.896 & 3.162 & 3.060 & 2.686 & 2.507 & 2.121 \\
6554 & 4.00 & 1.66 &  2.5             & 0.6     & 6.6     & 3.2   &  2.935 &  3.017 &  2.604 &  2.301 &  2.079  & 3.796 & 2.899 & 3.161 & 3.059 & 2.685 & 2.507 & 2.120 \\
6554 & 4.00 & 1.66 &  2.5             & 0.6     & 3.3     & 3.2   &  2.935 &  3.018 &  2.605 &  2.302 &  2.079  & 3.794 & 2.896 & 3.162 & 3.060 & 2.686 & 2.507 & 2.121 \\
6554 & 4.00 & 1.66 &  2.5             & 0.6     & 4.7     & 2.6   &  2.929 &  3.019 &  2.609 &  2.306 &  2.083  & 3.768 & 2.893 & 3.160 & 3.061 & 2.690 & 2.511 & 2.125 \\
6554 & 4.00 & 1.66 &  2.5             & 0.6     & 4.7     & 4.1   &  2.939 &  3.014 &  2.600 &  2.296 &  2.075  & 3.820 & 2.900 & 3.161 & 3.056 & 2.680 & 2.501 & 2.116 \\
\hline
\end{tabular}
\tablefoot{A stellar radius of 9.30 R$_{\odot}$ was assumed for Pollux, according to \citet{auriere15}. For Procyon, a radius of 2.05 R$_{\odot}$ was calculated from the effective temperature and luminosity of \citet{heiter15}.}
\end{center}
\end{sidewaystable*}

\begin{figure*}[t]
\centering
\includegraphics[width=0.47\textwidth]{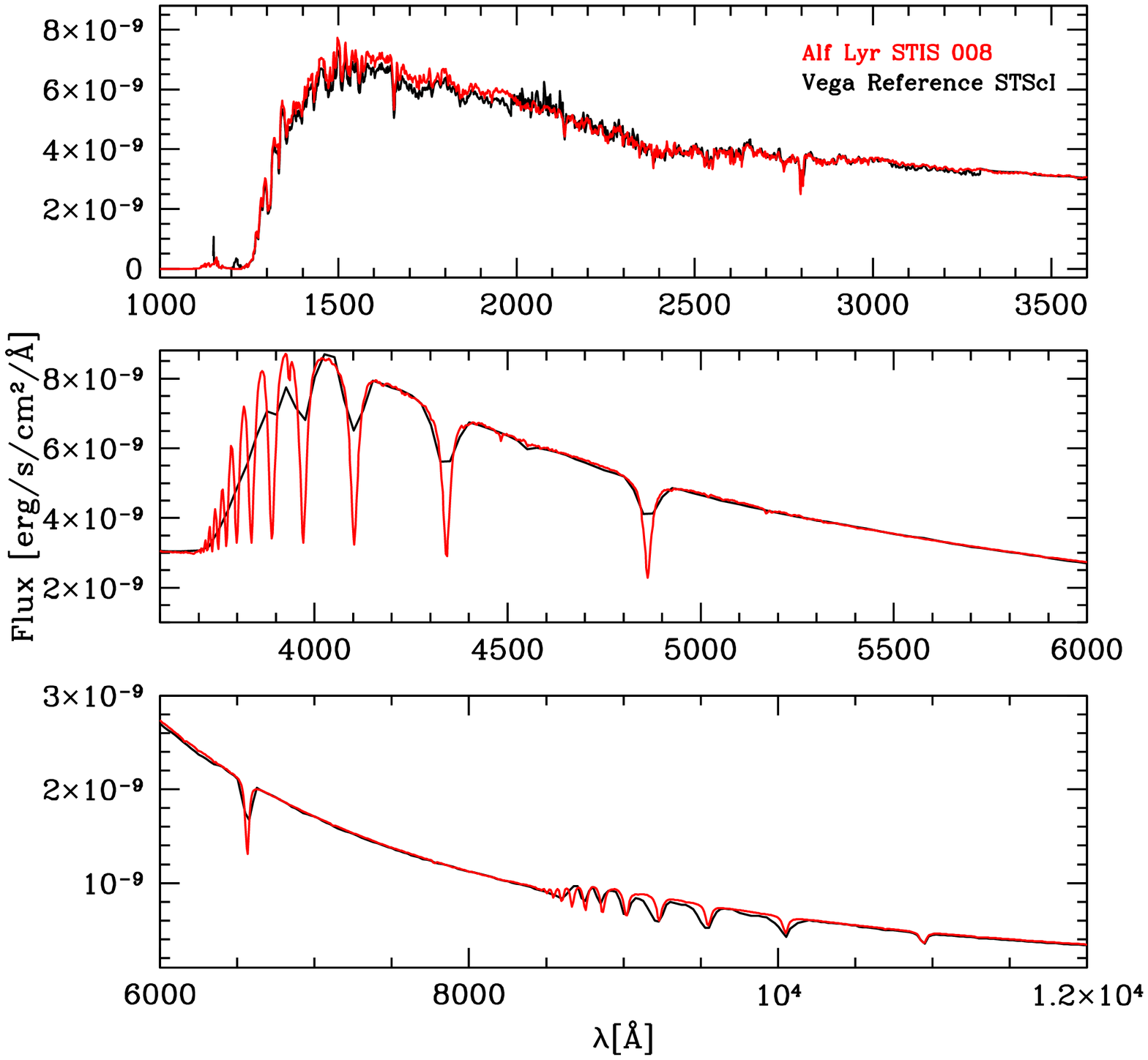}  
\includegraphics[width=0.47\textwidth]{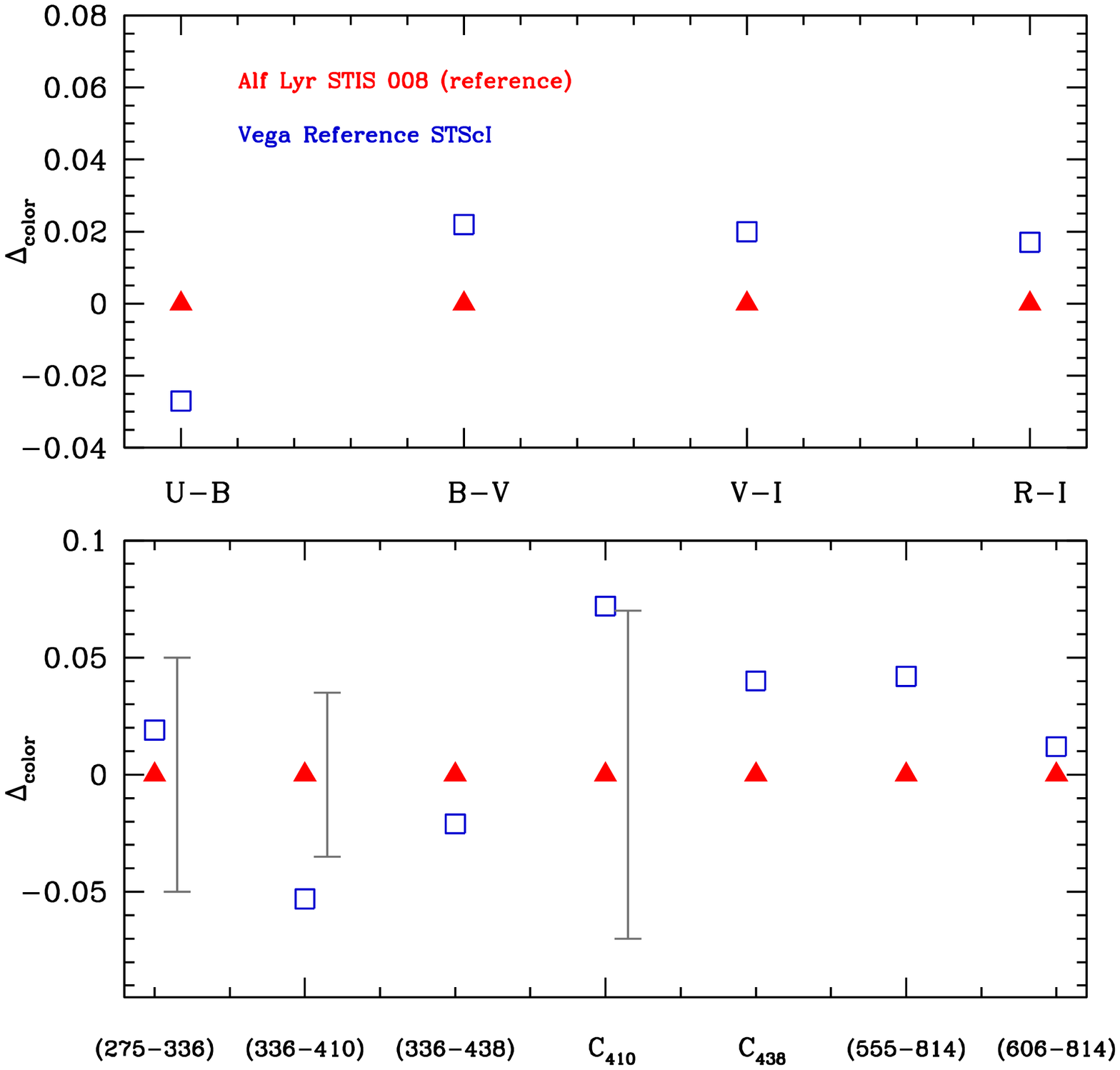}
\caption{\textit{Left}: Comparison between two Vega reference spectra. \textit{Right}: Difference in the magnitudes of Pollux caused by photometric calibrations based on the two Vega reference spectra. The gray vertical lines have the same meaning as in Fig.\ \ref{mag_pollux}.}
\label{mag_effectVega}
\end{figure*}

\begin{table}
\begin{center}
\caption{Dispersion in colors shown in Fig.\ \ref{mag_pollux} and \ref{mag_procyon}.}
\label{sig_colors}
\begin{tabular}{lccccccccccccccccc}
\hline
Color  & Pollux & Procyon\\    
\hline
($U-B$)       & 0.022 & 0.015 \\
($B-V$)       & 0.011 & 0.006 \\
($V-R$)       & 0.006 & 0.002 \\
($R-I$)       & 0.005 & 0.005 \\
(275-336) & 0.041 & 0.014 \\
(336-410) & 0.020 & 0.020 \\
(336-438) & 0.024 & 0.020 \\
C$_{410}$  & 0.031 & 0.021 \\
C$_{438}$  & 0.030 & 0.020 \\
(555-814) & 0.011 & 0.005 \\
(606-814) & 0.008 & 0.004 \\
\hline
\end{tabular}
\end{center}
\end{table}

\smallskip

In Fig.\ \ref{mag_procyon} we gather the color differences for Procyon. In Johnson photometry, the ($U-B$) color is the most affected by parameters variations (differences of up to 0.04 mag and a dispersion of 0.015). The smallest variation is observed in the ($V-R$) color, with a dispersion of 0.002 magnitude. For ($B-V$), the dispersion is 0.006. The color differences are smaller than in the case of Pollux. In the HST filters, colors involving filters located below 4500 \AA\ are most affected, with dispersion variations of up to 0.05 mag and dispersions reaching 0.02 mag. In these colors, the dispersion is smaller than the width of the TO in the globular cluster NGC~6752, but the range of colors can be of the same size. For colors based on filters covering redder parts of the spectrum, the dispersion drops to below the TO width.

%-------------------------------------------------------------------
\subsubsection{Photometric calibration: effect of the Vega reference spectrum}
\label{s_Vega}

Synthetic photometry requires calibration on a reference spectrum. In the VEGAMAG system, the star Vega is used for this. Its magnitude is set to 0.0 in all filters. In practice, this means that a correction factor (the zero point) must be applied to the integral of the stellar flux over the filter passband. Hence the final photometry depends on the choice of the Vega reference spectrum. In Fig.\ \ref{mag_effectVega} we show the difference in Johnson and HST photometry when using two different Vega reference spectra. The two spectra were retrieved from the HST calibration database\footnote{\url{ftp://ftp.stsci.edu/cdbs/current_calspec/}}. The ``Vega reference STScI'' spectrum was used by \citet{bedin05}. The spectrum ``Alf Lyr STIS 008'' is the spectrum currently used in the calibration of HST data. The difference between them is the use of the \citet{hayes85} Vega spectrum in the optical up to 1.05 $\mu$m and an ATLAS12 model (binned to a 25\AA\ resolution) beyond that limit for ``Vega reference STScI'' spectrum; the STIS spectrum from 1675 to 5350 \AA\ and an ATLAS12 model with \teff\ = 9400 K for the ``Alf Lyr STIS 008'' spectrum. The two spectra are compared in the left panel of Fig.~\ref{mag_effectVega}. Differences are present especially near the Balmer jump.

The right panel of Fig.\ \ref{mag_effectVega} illustrates the effect of changing the Vega reference spectrum on the photometry of Pollux. The differences are large, reaching 0.07 magnitudes in the C$_{410}$ color index. All colors are affected. It is therefore mandatory to treat the zero points consistently to compare observed to synthetic colors.

%-------------------------------------------------------------------
\subsubsection{Effect of extinction}
\label{s_effext}

Extinction affects the SED of stars differentially, being stronger at shorter wavelength. Extinction is characterized by two main quantities: the ratio of extinction at wavelength $\lambda$ compared to that at a reference wavelength (usually in the $V$ or $K$ band), this is the extinction law; and the total extinction at the reference wavelength. To quantify the effect of extinction on synthetic photometry, we have used two sets of extinction laws. The first is a combination of the extinction law of \citet{seaton79} in the ultraviolet and of \citet{howarth83} in the optical. The second is the extinction law of \citet{ccm89}.
We have parameterized the total extinction by A$_V = R_V \times\ E(B-V),$ where $R_V$ is the ratio of total to selective extinction, which we held fixed to 3.2, and $E(B-V) = (B-V) - (B-V)_0$ with $(B-V)_0$ the intrinsic color. 

\begin{figure}[t]
\centering
\includegraphics[width=9cm]{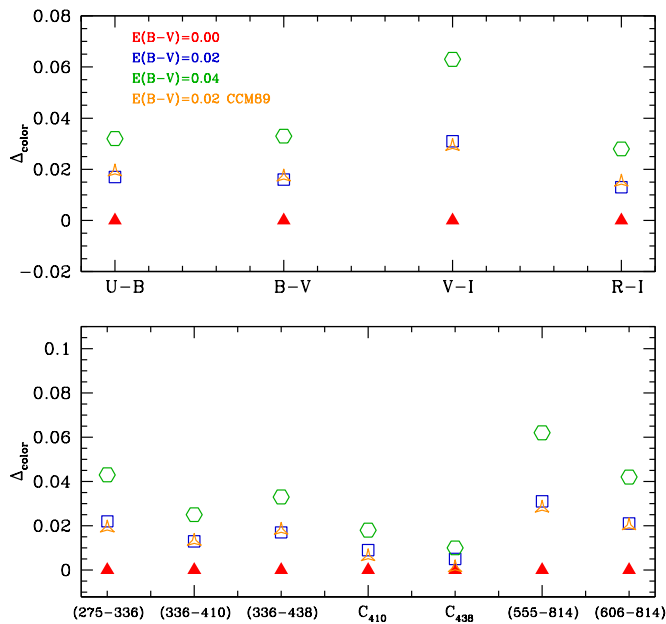}
\caption{Effect of extinction on colors for Pollux. Red triangles, blue squares, and green hexagons correspond to the extinction laws of  \citet{seaton79} in the ultraviolet and of \citet{howarth83} in the optical. The orange empty triangles refer to calculations made with the extinction law of \citet{ccm89}. $\Delta_{color}$ is the color difference relative to the models shown by red triangles.}
\label{mag_ext}
\end{figure}

Fig.\ \ref{mag_ext} shows the effect of extinction on synthetic colors. A variation of 0.02 in E($B-V$) translates into variations between 0.015 and 0.10 in colors depending on the filters used. The changes are largest for colors based on filters that are
more separated in wavelength. For a given observed ($B-V$), a variation of 0.02 in E($B-V$) corresponds to an uncertainty of 0.02 in intrinsic ($B-V$)$_0$. For comparison, a K0~III star (spectral type of Pollux) has ($B-V$)$_0$=0.81, while a K1~III star has ($B-V$)$_0$=0.86 \citep{lang93}, or a difference in intrinsic color of 0.05. Hence our test corresponds to an error smaller than one spectral sub-type in spectral classification. The choice of extinction law also affects the resulting colors, the difference between our two laws being $<$ 0.01.

%-------------------------------------------------------------------
\subsubsection{Effect of atmospheric correction}
\label{s_airmass}

\begin{figure}[t]
\centering
\includegraphics[width=9cm]{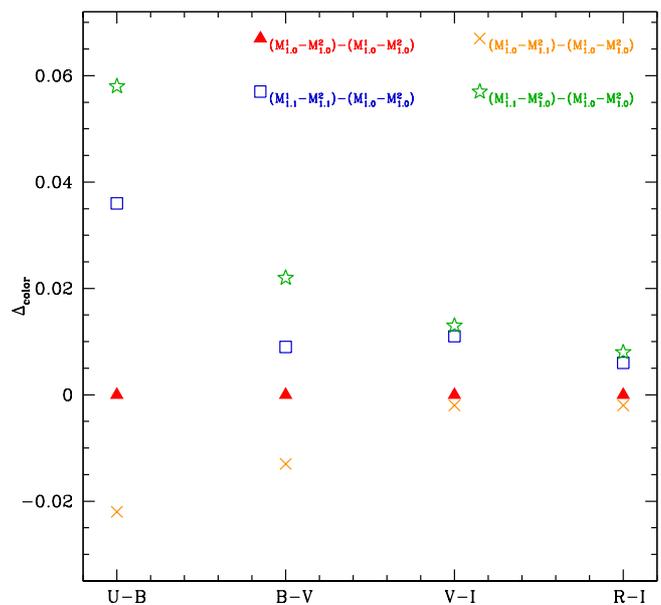}
\caption{Effect of atmospheric correction on colors based on $UBVRI$ photometry. $M^1$ and $M^2$ are the first and second magnitude used to build a given color (e.g., $M^1$=$U$ and $M^2$=$B$ for color ($U-B$)). The subscripts (1.0 and 1.1) correspond to the airmass adopted for the computation of the atmospheric corrections. $\Delta_{color}$ is the color difference relative to the models shown by red triangles.}
\label{mag_airmass}
\end{figure}

For ground-based observations a correction for the absorption in the Earth's atmosphere has to be performed. The absorption is stronger at shorter wavelength and increases with airmass. In our calculations, we adopted the correction coefficient for the ESO/La Silla observatory provided by \citet{burki95}. Fig.\ \ref{mag_airmass} shows the effect of airmass on colors based on $UBVRI$ photometry. As expected, colors are bluer when the airmass increases from 1.0 to 1.1. The difference remains below 0.01 magnitude when the $U$ filter is not used. For ($U-B$), the airmass increase leads to a color 0.035 magnitudes bluer. The stars and crosses correspond to a case where photometry was acquired in two different airmass conditions for the filters used in a given color. In this configuration, color differences can reach almost 0.06 magnitude in ($U-B$). They remain below 0.02 magnitude for the other colors.

%%%%%%%%%%%%%%%%%%%%%%%%%%%%%%%%%%%%%%%%%%%%%%%%%%%%%%%%%%%%%%
%
\subsection{SED fit}
\label{s_sed}

So far, we have assumed that theoretical spectra perfectly reproduce the SED of Pollux and Procyon. In this section we investigate to which degree this is correct.

\begin{figure}[]
\centering
\includegraphics[width=9cm]{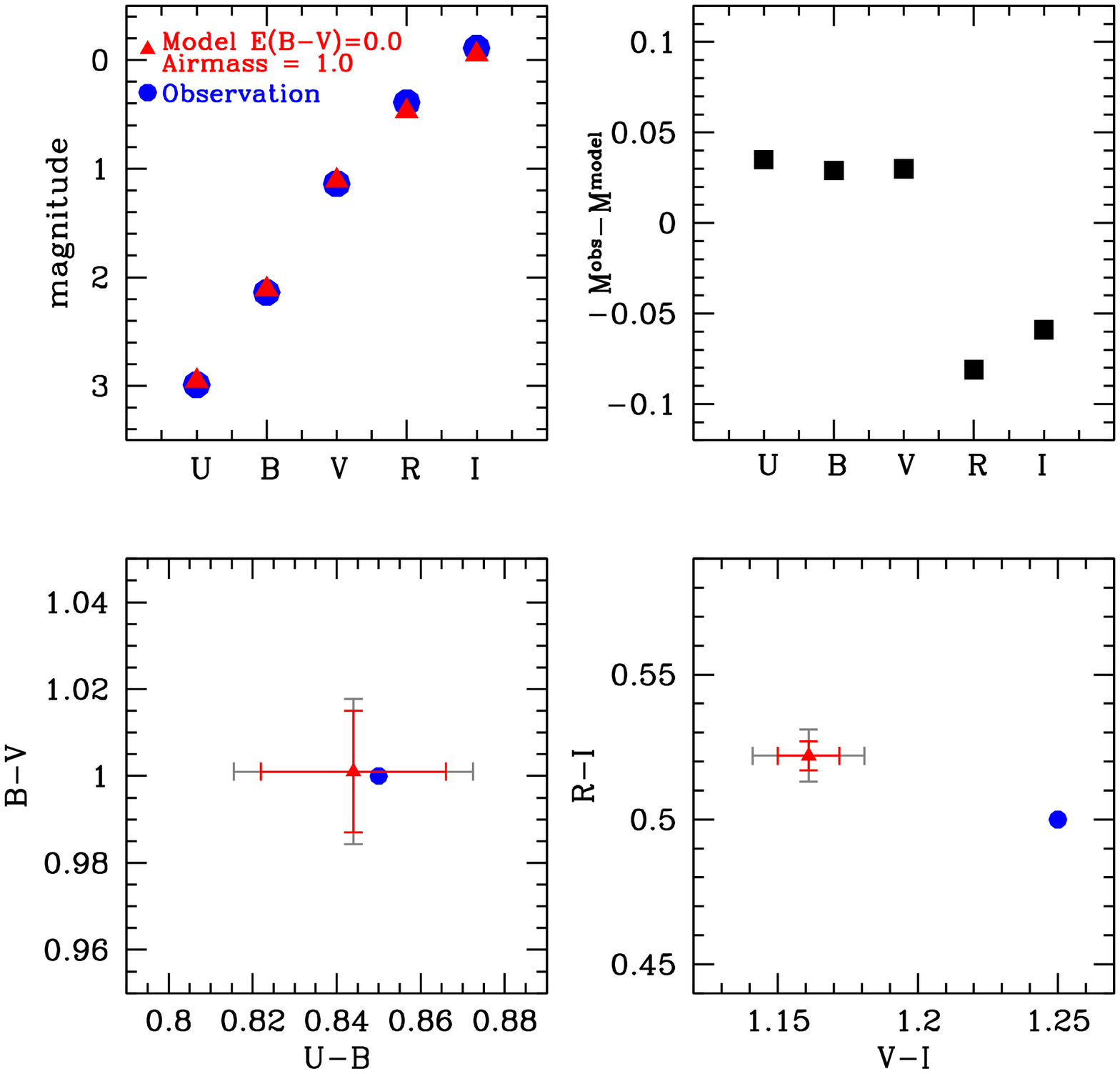}
\caption{Comparison of observed magnitudes and colors of Pollux (blue circles) with those predicted by the reference model (red triangles). An airmass of 1.0, a radius of 9.3 R$_{\odot}$ and a distance of 10.36 pc are assumed. Solid error bars take into account only uncertainties due to stellar parameters (Table\ \ref{sig_colors}). Gray error bars take into account an additional contribution due to extinction and airmass. The former is set to half the difference in colors between models with E($B-V$)=0.00 and E($B-V$)=0.02, the latter to half the difference between corrections for an airmass of 1.0 and 1.1. In the upper right panel, the black squares show the difference between the observed and predicted magnitudes for the five Johnson filters.}
\label{mag_pol}
\end{figure}

Fig.\ \ref{mag_pol} shows the ground-based $UBVRI$ photometry of Pollux according to \citet{ducati02}. It is identical to that of the GCPD database from which we have retrieved the filters throughputs. We added the magnitudes computed from our reference model, together with error bars adopted from Sect.\ \ref{s_unc}. Our model reproduces the $UBV$ photometry very well, but faces problems with the $R$ and $I$ filters. From the top and bottom right panels, it appears that the model lacks flux in both bands, which translates into a too blue $V-I$ color and a too red $R-I$ color. The problems are most severe in $V-I,$ where the mismatch between model and observations reaches 0.10 mag.

\begin{figure*}[]
\centering
\includegraphics[width=0.49\textwidth]{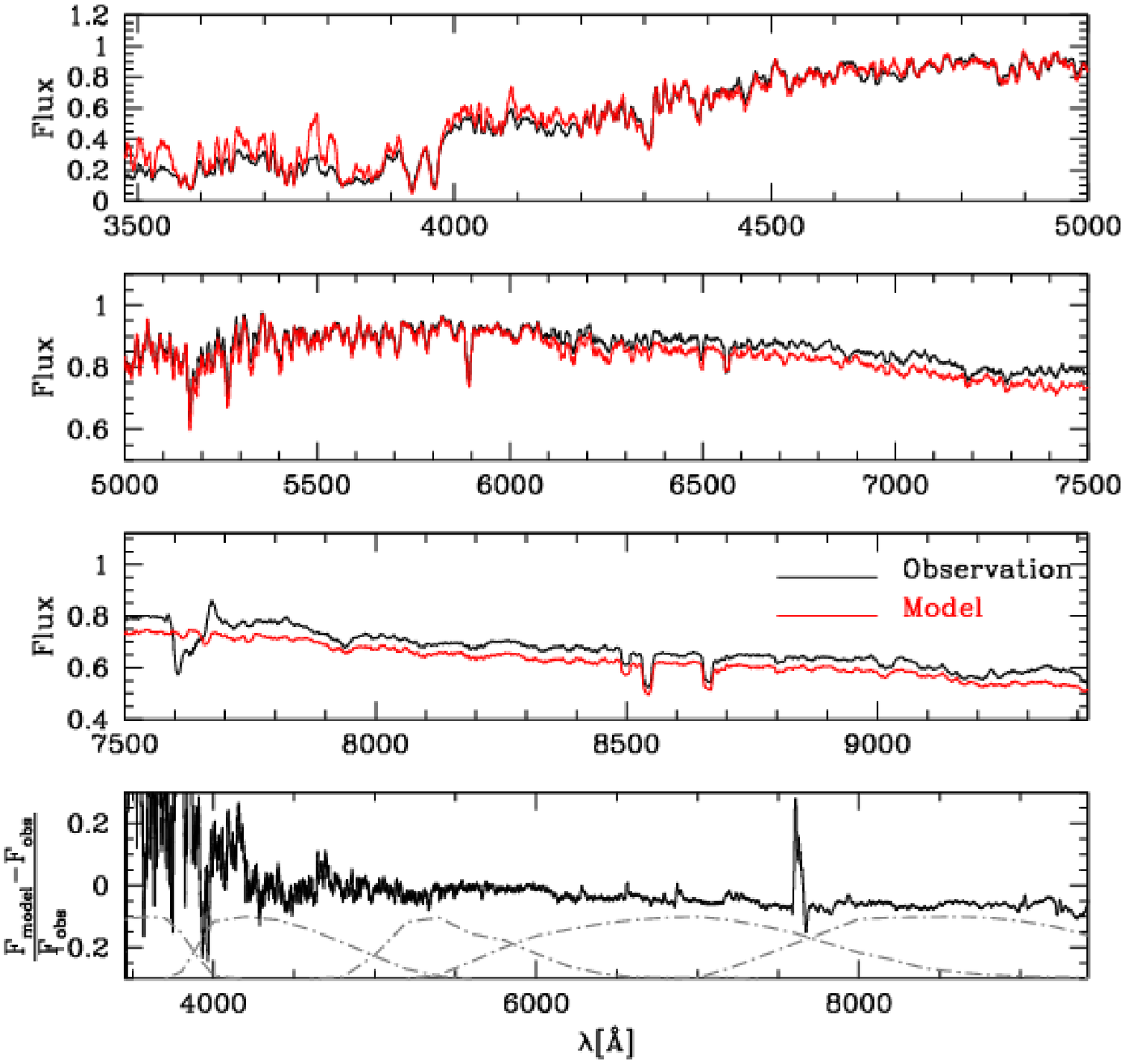}
\includegraphics[width=0.49\textwidth]{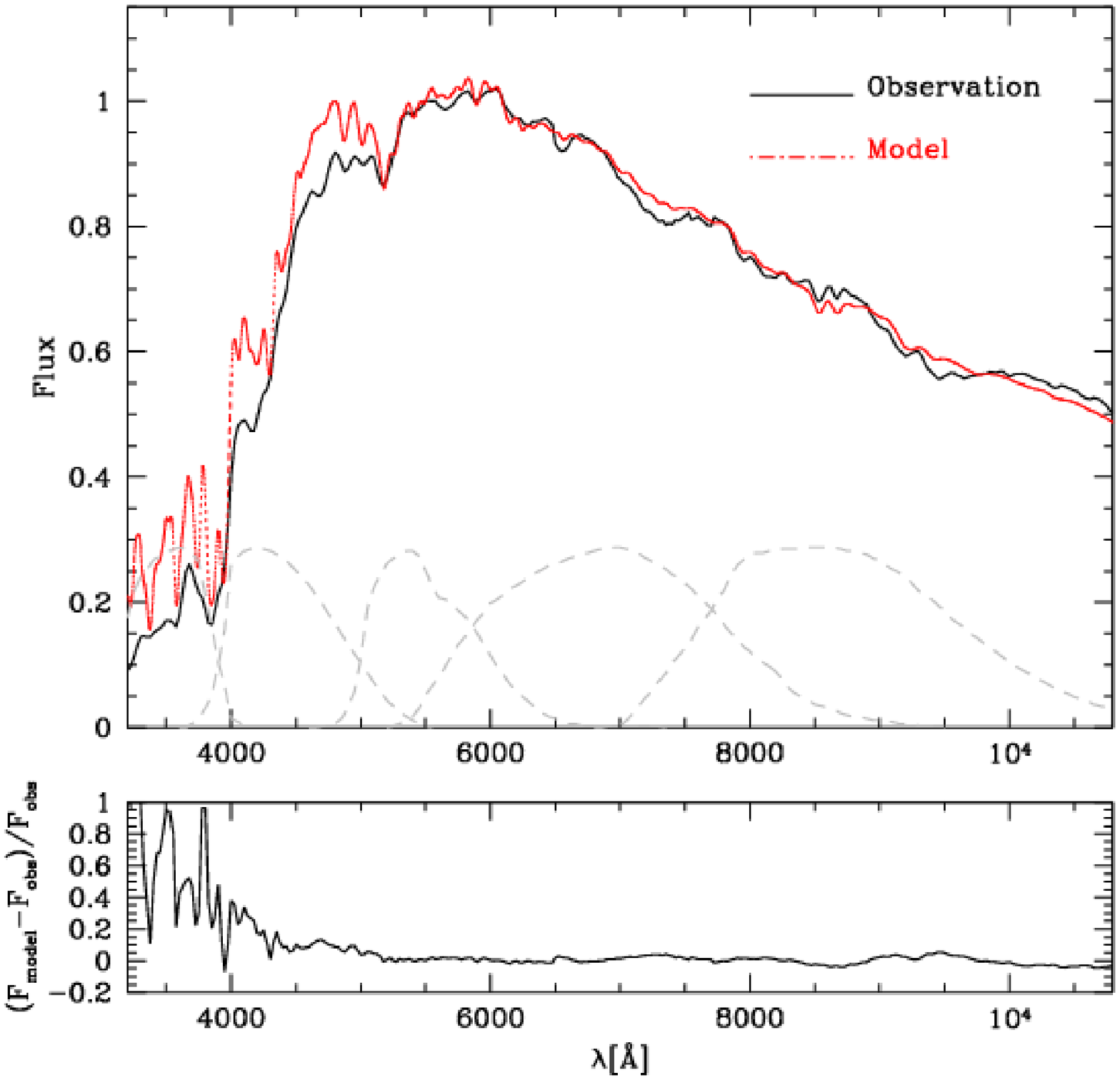}
\caption{Comparison between the Pollux model computed with the parameters obtained from spectroscopy (red line) and the observed spectrum of \citet{valdes04} (left panel) / the SED of \citet{alek97} (right panel). The model was degraded to the resolution of the observed spectra (R$\sim$10000 in the left panel, R$\sim$100 in the right panel). In addition, the model and the observed spectrum of the left panel were smoothed for clarity of the comparison. In both panels the spectra have been normalized with respect to their flux at 5500 \AA. The dot-dashed line shows the $UBVRI$ filters throughputs. The bottom panels show the difference between model and observation.}
\label{fit_sed_pollux}
\end{figure*}

Fig.\ \ref{fit_sed_pollux} shows the comparison of the reference model and two spectra observed from the ground: the medium-resolution spectrum of \citet{valdes04} (left panel), and the low-resolution spectrum of \citet{alek97}. The agreement between the model and the observed spectrum is good. Differences between magnitudes calculated from the spectra presented in Fig.\ \ref{fit_sed_pollux} are shown in Table \ref{col_diff}. The $V$ and $R$ magnitudes are very similar between the synthetic and the observed spectra (within 0.04 magnitudes), regardless of the observed spectrum. In the B band, differences vary from 0.02 to 0.13 magnitude depending on the observed spectrum. This presumably shows that flux calibration is critical since the two observed spectra do not show the same flux level in the blue, while Pollux is supposed to have a stable flux level (see Sect.\ \ref{s_targets}). The $U$ and $I$ bands are almost fully probed only by the spectrum of \citet{alek97}. We calculated the corresponding magnitudes on the wavelength range covered by this spectrum (i.e., we cut the synthetic spectrum below and above the limits of the observed spectrum). The $I$ band is very well reproduced by our model, while a difference of 0.50 magnitude appears in the $U$ band. These results indicate that the ($V-I$) color of our model reproduces (within less than 0.02 magnitude) the ($V-I$) color that is obtained from the spectrum of \citet{alek97} well, while there is a mismatch in ($V-I$) in Fig.\ \ref{mag_pol}. Alternatively, the ($U-B$) color of our model reproduces the observed color in Fig.\ \ref{mag_pol} very
well, while the spectrum of \citet{alek97} has much less flux than our model in the $U$ band. 

\smallskip

\begin{figure}[]
\centering
\includegraphics[width=9cm]{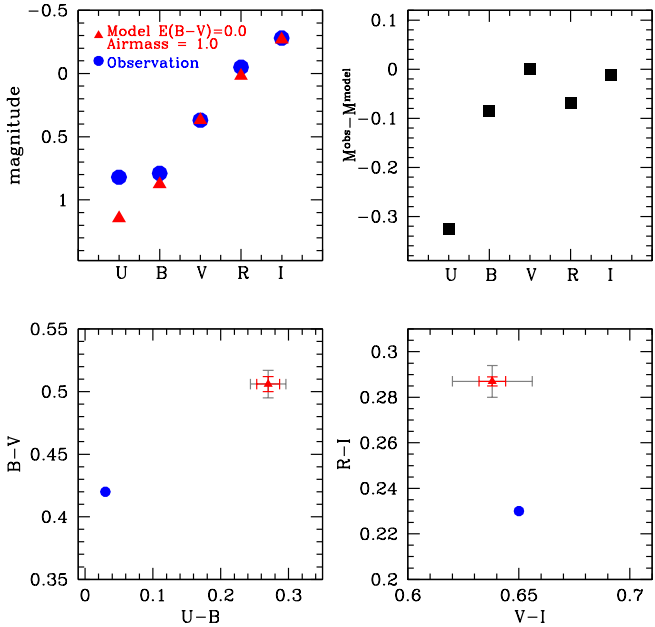}
\caption{Same as Fig.\ \ref{mag_pol} for Procyon. The radius was adjusted so that the $V$ magnitude from the model matches the observed magnitude.}
\label{mag_pro}
\end{figure}

\begin{table}
\begin{center}
\caption{Difference between magnitudes calculated from the synthetic and observed spectrum of Pollux and Procyon.}
\label{col_diff}
\begin{tabular}{lccccc}
\hline
Observed spectrum &   $\Delta U$  & $\Delta B$   & $\Delta V$   & $\Delta R$    &  $\Delta I$   \\    
\hline
 & & Pollux & \\
\hline
Valdes    & -- & 0.02 & 0.02 & 0.04 & --\\
Alekseeva & 0.50 & 0.13 & 0.02 & 0.01 & 0.01 \\
\hline
 & &  Procyon & \\
\hline
Alekseeva & 0.11 & 0.03 & 0.00 & 0.00 & 0.00\\
\hline
\end{tabular}
\end{center}
\end{table}

\begin{figure*}[]
\centering
\includegraphics[width=0.49\textwidth]{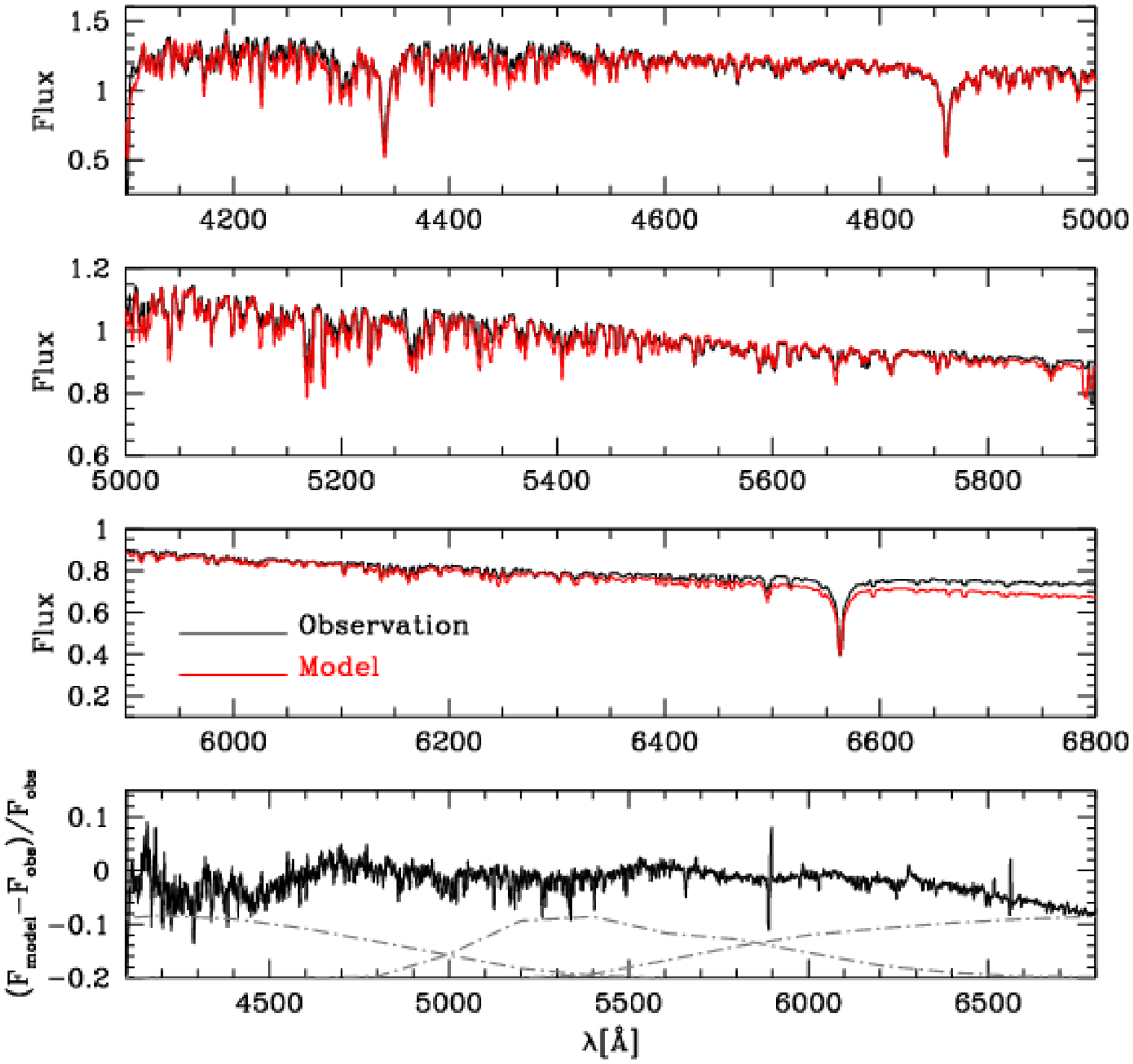}
\includegraphics[width=0.49\textwidth]{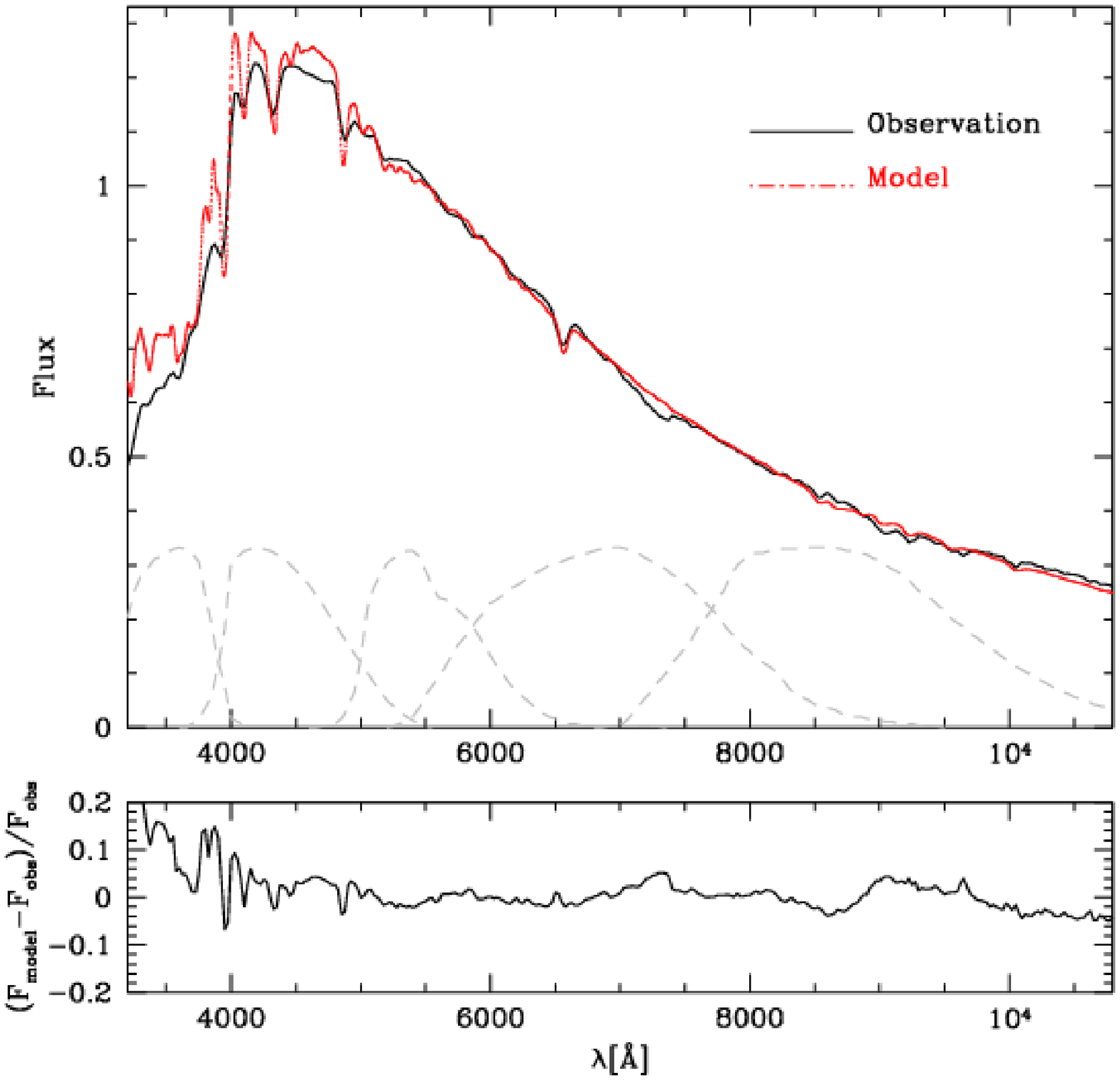}
\caption{Same as Fig.\ \ref{fit_sed_pollux} for Procyon. The left (right) panel shows the ELODIE spectrum of \citet{prusou01} \citep{alek97}.}
\label{fit_sed_procyon}
\end{figure*}

Ground-based $UBVRI$ photometry from our reference model of Procyon is compared to observed photometry in Fig.\ \ref{mag_pro}. A comparison of the Procyon ground-based spectrum of \citet{alek97} and \citet{prusou01} with our model is shown in Fig.\ \ref{fit_sed_procyon}\footnote{There is no Procyon spectrum in the database of \citet{valdes04}.}. From this figure and Table \ref{col_diff}, we conclude that the model reproduces the observed spectrum in the $VRI$ bands very
well and that deviations appear in the $B$ and mostly $U$ band. Fig.\ \ref{mag_pro} confirms that the synthetic colors involving the $U$ and $B$ band are problematic. However, the observed $U$ magnitude indicates a higher flux than predicted, while the opposite is seen in Fig.\ \ref{fit_sed_procyon}, where the spectrum of \citet{alek97} has less flux than our model shortward of 4000 \AA. Fig.\ \ref{mag_pro} shows that the the theoretical ($R-I$) color is 0.06 magnitude redder than the color obtained from imaging. This trend is not confirmed by the direct comparison of the Alekseeva et al.\ spectrum in Fig.\ \ref{fit_sed_procyon}: according to Table \ref{col_diff}, the $R$ and $I$ magnitudes calculated from the Alekseeva spectrum are the same as those of our Procyon reference model, hence ($R-I$) is also the same.

\smallskip

Our comparisons indicate that model reasonably well reproduces flux-calibrated observed spectra. When we compare photometry computed from the synthetic spectra to photometry resulting from imaging, discrepancies appear. Given the uncertainties in photometry based on filters with passbands covering (part of) the wavelength range below $\sim$4500 \AA, this is expected for $U$ and $B$ filters. The companion white dwarf to Procyon may explain part of the discrepant ($U-B$) and ($B-V$) colors. However, the mismatch observed for $R$ and $I$ filters is worrisome. The magnitude of the discrepancy between observed and synthetic ($V-I$) (or
($R-I$)) for Pollux (for Procyon) cannot be attributed to incorrect modeling of the spectra of these stars since comparisons to observed SEDs are quantitatively rather good. We speculate that difference between the calibration process of our synthetic photometry and the reduction and calibration of the observed photometry is responsible for the mismatch. 
This stresses the need for accurate calibrations and for the publication of all the reduction details in order to minimize systematic errors. This is crucial for performing synthetic photometry at the level of 0.01 mag accuracy, a level required if blue filters
are to be used,
which are best suited to studying multiple populations in globular clusters.

%%%%%%%%%%%%%%%%%%%%%%%%%%%%%%%%%%%%%%%%%%%%%%%%%%%%%%%%%%%%%%%%%%%%%%%%%%%%%%%%%%%%%%%%%%%%%%%%%%%%%%%%%%%%%%%%%%%%%%%%%%%%%%%
%%%%%%%%%%%%%%%%%%%%%%%%%%%%%%%%%%%%%%%%%%%%%%%%%%%%%%%%%%%%%%%%%%%%%%%%%%%%%%%%%%%%%%%%%%%%%%%%%%%%%%%%%%%%%%%%%%%%%%%%%%%%%%%
\section{Conclusion and future work}
\label{s_conc}

We have presented a study of uncertainties on synthetic photometry in the context of the understanding the properties of globular clusters. Our goal was to provide an estimate of the dispersion that can be used to build artificial populations of stars centered on a theoretical isochrone. Such artificial populations can then
be compared to observed populations in CMDs to infer properties of globular clusters.

We have calculated atmosphere models and synthetic spectra with the codes ATLAS12 and SYNTHE, respectively. We chose two reference stars: Pollux, a K0III star typical of giants at the bottom of the RGB in globular clusters, and Procyon, an F5IV-V dwarf typical of TO stars. Using the best spectroscopic parameters and their uncertainties for these two stars, we studied the effect of effective temperature, surface gravity, microturbulent velocity, C, N, O, and Fe abundances on the resulting photometry. We also estimated the changes in photometry caused by uncertain extinction, by the airmass conditions, and by different calibrations of zero points in the VEGAMAG system.

We provide estimates of the dispersion to be expected in photometry based on $UBVRI$ and the following HST filters: F275W, F336W, F410M, F438W, F555W, F606W, and F814W. We show that uncertainties are larger at shorter wavelength, as was known before. Our results indicate that even if synthetic spectra reproduce  flux-calibrated SEDs well, synthetic photometry may not reproduce published $UBVRI$ photometry. This most likely reflects different reduction and
calibration processes and calls for the publication of all the details of such processes. This is crucial if a 0.01 mag accuracy, which is necessary to study the properties of multiple populations in globular clusters, is to be reached by synthetic photometry. 

Regardless of these issues, the effects of uncertain stellar and observational parameters on synthetic colors will be used in subsequent studies to produce synthetic CMDs that include a realistic treatment of errors. In practice, artificial populations will be built from theoretical isochrones and the dispersion estimated in the present study. The ability of theoretical isochrones to reproduce the location of multiple populations in globular clusters will be tested. This will be useful to constrain the physics of evolutionary models providing isochrones, the physics of atmosphere models that provide synthetic photometry, and ultimately, it will bring additional constraints to some properties of globular clusters (helium content and age).

%%#####################################################################
\begin{acknowledgements}
We thank an anonymous referee for comments that helped to clarify the goal of this study.
We thank Fiorella Castelli, Robert Kurucz, and Marwan Gebran for help with the codes ATLAS12 and SYNTHE. We thank Corinne Charbonnel and William Chantereau for fruitful discussions. We warmly thank Antonino Milone for sharing his HST photometry of NGC~6752.
This research has made use of the SVO Filter Profile Service supported from the Spanish MINECO through grant AyA2014-55216.
This research has made use of the SIMBAD database, operated at CDS, Strasbourg, France.
We thank the french ``Programme National de Physique Stellaire (PNPS)'' of CNRS/INSU for financial support.
\end{acknowledgements}
%%#####################################################################
\bibliographystyle{aa}
\bibliography{article}

\begin{thebibliography}{69}
\expandafter\ifx\csname natexlab\endcsname\relax\def\natexlab#1{#1}\fi

\bibitem[{{Adamczak} \& {Lambert}(2013)}]{adam13}
{Adamczak}, J. \& {Lambert}, D.~L. 2013, \apj, 765, 155

\bibitem[{{Alekseeva} {et~al.}(1997){Alekseeva}, {Arkharov}, {Galkin},
  {Hagen-Thorn}, {Nikanorova}, {Novikov}, {Novopashenny}, {Pakhomov}, {Ruban},
  \& {Shechegolev}}]{alek97}
{Alekseeva}, G.~A., {Arkharov}, A.~A., {Galkin}, V.~D., {et~al.} 1997, Baltic
  Astronomy, 6, 481

\bibitem[{{Auri{\`e}re} {et~al.}(2015){Auri{\`e}re}, {Konstantinova-Antova},
  {Charbonnel}, {Wade}, {Tsvetkova}, {Petit}, {Dintrans}, {Drake}, {Decressin},
  {Lagarde}, {Donati}, {Roudier}, {Ligni{\`e}res}, {Schr{\"o}der},
  {Landstreet}, {L{\`e}bre}, {Weiss}, \& {Zahn}}]{auriere15}
{Auri{\`e}re}, M., {Konstantinova-Antova}, R., {Charbonnel}, C., {et~al.} 2015,
  \aap, 574, A90

\bibitem[{{Bedin} {et~al.}(2005){Bedin}, {Cassisi}, {Castelli}, {Piotto},
  {Anderson}, {Salaris}, {Momany}, \& {Pietrinferni}}]{bedin05}
{Bedin}, L.~R., {Cassisi}, S., {Castelli}, F., {et~al.} 2005, \mnras, 357, 1038

\bibitem[{{Bedin} {et~al.}(2004){Bedin}, {Piotto}, {Anderson}, {Cassisi},
  {King}, {Momany}, \& {Carraro}}]{bedin04}
{Bedin}, L.~R., {Piotto}, G., {Anderson}, J., {et~al.} 2004, \apjl, 605, L125

\bibitem[{{Burki} {et~al.}(1995){Burki}, {Rufener}, {Burnet}, {Richard},
  {Blecha}, \& {Bratschi}}]{burki95}
{Burki}, G., {Rufener}, F., {Burnet}, M., {et~al.} 1995, \aaps, 112, 383

\bibitem[{{Cardelli} {et~al.}(1989){Cardelli}, {Clayton}, \& {Mathis}}]{ccm89}
{Cardelli}, J.~A., {Clayton}, G.~C., \& {Mathis}, J.~S. 1989, \apj, 345, 245

\bibitem[{{Carretta}(2013)}]{car13}
{Carretta}, E. 2013, \aap, 557, A128

\bibitem[{{Carretta}(2015)}]{car15}
{Carretta}, E. 2015, \apj, 810, 148

\bibitem[{{Carretta} {et~al.}(2009{\natexlab{a}}){Carretta}, {Bragaglia},
  {Gratton}, {D'Orazi}, \& {Lucatello}}]{car09fe}
{Carretta}, E., {Bragaglia}, A., {Gratton}, R., {D'Orazi}, V., \& {Lucatello},
  S. 2009{\natexlab{a}}, \aap, 508, 695

\bibitem[{{Carretta} {et~al.}(2006){Carretta}, {Bragaglia}, {Gratton}, {Leone},
  {Recio-Blanco}, \& {Lucatello}}]{car06}
{Carretta}, E., {Bragaglia}, A., {Gratton}, R.~G., {et~al.} 2006, \aap, 450,
  523

\bibitem[{{Carretta} {et~al.}(2010){Carretta}, {Bragaglia}, {Gratton},
  {Lucatello}, {Bellazzini}, {Catanzaro}, {Leone}, {Momany}, {Piotto}, \&
  {D'Orazi}}]{car10}
{Carretta}, E., {Bragaglia}, A., {Gratton}, R.~G., {et~al.} 2010, \aap, 520,
  A95

\bibitem[{{Carretta} {et~al.}(2009{\natexlab{b}}){Carretta}, {Bragaglia},
  {Gratton}, {Lucatello}, {Catanzaro}, {Leone}, {Bellazzini}, {Claudi},
  {D'Orazi}, {Momany}, {Ortolani}, {Pancino}, {Piotto}, {Recio-Blanco}, \&
  {Sabbi}}]{car09}
{Carretta}, E., {Bragaglia}, A., {Gratton}, R.~G., {et~al.} 2009{\natexlab{b}},
  \aap, 505, 117

\bibitem[{{Carretta} {et~al.}(2012){Carretta}, {Bragaglia}, {Gratton},
  {Lucatello}, \& {D'Orazi}}]{car12}
{Carretta}, E., {Bragaglia}, A., {Gratton}, R.~G., {Lucatello}, S., \&
  {D'Orazi}, V. 2012, \apjl, 750, L14

\bibitem[{{Carretta} {et~al.}(2007){Carretta}, {Bragaglia}, {Gratton},
  {Lucatello}, \& {Momany}}]{car07}
{Carretta}, E., {Bragaglia}, A., {Gratton}, R.~G., {Lucatello}, S., \&
  {Momany}, Y. 2007, \aap, 464, 927

\bibitem[{{Carretta} {et~al.}(2005){Carretta}, {Gratton}, {Lucatello},
  {Bragaglia}, \& {Bonifacio}}]{car05}
{Carretta}, E., {Gratton}, R.~G., {Lucatello}, S., {Bragaglia}, A., \&
  {Bonifacio}, P. 2005, \aap, 433, 597

\bibitem[{{Cassisi} {et~al.}(2017){Cassisi}, {Salaris}, {Pietrinferni}, \&
  {Hyder}}]{cas17}
{Cassisi}, S., {Salaris}, M., {Pietrinferni}, A., \& {Hyder}, D. 2017, \mnras,
  464, 2341

\bibitem[{{Chantereau} {et~al.}(2016){Chantereau}, {Charbonnel}, \&
  {Meynet}}]{chantereau16}
{Chantereau}, W., {Charbonnel}, C., \& {Meynet}, G. 2016, \aap, 592, A111

\bibitem[{{Charbonnel} {et~al.}(2013){Charbonnel}, {Chantereau}, {Decressin},
  {Meynet}, \& {Schaerer}}]{charb13}
{Charbonnel}, C., {Chantereau}, W., {Decressin}, T., {Meynet}, G., \&
  {Schaerer}, D. 2013, \aap, 557, L17

\bibitem[{{Decressin} {et~al.}(2007){Decressin}, {Meynet}, {Charbonnel},
  {Prantzos}, \& {Ekstr{\"o}m}}]{dec07}
{Decressin}, T., {Meynet}, G., {Charbonnel}, C., {Prantzos}, N., \&
  {Ekstr{\"o}m}, S. 2007, \aap, 464, 1029

\bibitem[{{Denissenkov} \& {Hartwick}(2014)}]{dh14}
{Denissenkov}, P.~A. \& {Hartwick}, F.~D.~A. 2014, \mnras, 437, L21

\bibitem[{{Dotter} {et~al.}(2015){Dotter}, {Ferguson}, {Conroy}, {Milone},
  {Marino}, \& {Yong}}]{dotter15}
{Dotter}, A., {Ferguson}, J.~W., {Conroy}, C., {et~al.} 2015, \mnras, 446, 1641

\bibitem[{{Ducati}(2002)}]{ducati02}
{Ducati}, J.~R. 2002, VizieR Online Data Catalog, 2237

\bibitem[{{Gieles} {et~al.}(2018){Gieles}, {Charbonnel}, {Krause},
  {H{\'e}nault-Brunet}, {Agertz}, {Lamers}, {Bastian}, {Gualandris}, {Zocchi},
  \& {Petts}}]{gieles18}
{Gieles}, M., {Charbonnel}, C., {Krause}, M.~G.~H., {et~al.} 2018, \mnras
  [\eprint[arXiv]{1804.04682}]

\bibitem[{{Gratton} {et~al.}(2007){Gratton}, {Lucatello}, {Bragaglia},
  {Carretta}, {Cassisi}, {Momany}, {Pancino}, {Valenti}, {Caloi}, {Claudi},
  {D'Antona}, {Desidera}, {Fran{\c c}ois}, {James}, {Moehler}, {Ortolani},
  {Pasquini}, {Piotto}, \& {Recio-Blanco}}]{gratton07}
{Gratton}, R.~G., {Lucatello}, S., {Bragaglia}, A., {et~al.} 2007, \aap, 464,
  953

\bibitem[{{Gray}(2014)}]{gray14}
{Gray}, D.~F. 2014, \apj, 796, 88

\bibitem[{{Gruyters} {et~al.}(2017){Gruyters}, {Casagrande}, {Milone},
  {Hodgkin}, {Serenelli}, \& {Feltzing}}]{gruy17}
{Gruyters}, P., {Casagrande}, L., {Milone}, A.~P., {et~al.} 2017, \aap, 603,
  A37

\bibitem[{{Han} {et~al.}(2009){Han}, {Lee}, {Joo}, {Sohn}, {Yoon}, {Kim}, \&
  {Lee}}]{han09}
{Han}, S.-I., {Lee}, Y.-W., {Joo}, S.-J., {et~al.} 2009, \apjl, 707, L190

\bibitem[{{Hayes}(1985)}]{hayes85}
{Hayes}, D.~S. 1985, in IAU Symposium, Vol. 111, Calibration of Fundamental
  Stellar Quantities, ed. D.~S. {Hayes}, L.~E. {Pasinetti}, \& A.~G.~D.
  {Philip}, 225--249

\bibitem[{{Heiter} {et~al.}(2015){Heiter}, {Jofr{\'e}}, {Gustafsson}, {Korn},
  {Soubiran}, \& {Th{\'e}venin}}]{heiter15}
{Heiter}, U., {Jofr{\'e}}, P., {Gustafsson}, B., {et~al.} 2015, \aap, 582, A49

\bibitem[{{Howarth}(1983)}]{howarth83}
{Howarth}, I.~D. 1983, \mnras, 203, 301

\bibitem[{{Jofr{\'e}} {et~al.}(2015{\natexlab{a}}){Jofr{\'e}}, {Petrucci},
  {Saffe}, {Saker}, {de la Villarmois}, {Chavero}, {G{\'o}mez}, \&
  {Mauas}}]{jofre15b}
{Jofr{\'e}}, E., {Petrucci}, R., {Saffe}, C., {et~al.} 2015{\natexlab{a}},
  \aap, 574, A50

\bibitem[{{Jofr{\'e}} {et~al.}(2015{\natexlab{b}}){Jofr{\'e}}, {Heiter},
  {Soubiran}, {Blanco-Cuaresma}, {Masseron}, {Nordlander}, {Chemin}, {Worley},
  {Van Eck}, {Hourihane}, {Gilmore}, {Adibekyan}, {Bergemann}, {Cantat-Gaudin},
  {Delgado-Mena}, {Gonz{\'a}lez Hern{\'a}ndez}, {Guiglion}, {Lardo}, {de
  Laverny}, {Lind}, {Magrini}, {Mikolaitis}, {Montes}, {Pancino},
  {Recio-Blanco}, {Sordo}, {Sousa}, {Tabernero}, \& {Vallenari}}]{jofre15a}
{Jofr{\'e}}, P., {Heiter}, U., {Soubiran}, C., {et~al.} 2015{\natexlab{b}},
  \aap, 582, A81

\bibitem[{{Kraft} {et~al.}(1997){Kraft}, {Sneden}, {Smith}, {Shetrone},
  {Langer}, \& {Pilachowski}}]{kraft97}
{Kraft}, R.~P., {Sneden}, C., {Smith}, G.~H., {et~al.} 1997, \aj, 113, 279

\bibitem[{{Kravtsov} {et~al.}(2014){Kravtsov}, {Alca{\'{\i}}no}, {Marconi}, \&
  {Alvarado}}]{krav14}
{Kravtsov}, V., {Alca{\'{\i}}no}, G., {Marconi}, G., \& {Alvarado}, F. 2014,
  \apj, 783, 56

\bibitem[{{Kurucz}(2005)}]{kur05}
{Kurucz}, R.~L. 2005, Memorie della Societa Astronomica Italiana Supplementi,
  8, 76

\bibitem[{{Kurucz}(2014)}]{kur14}
{Kurucz}, R.~L. 2014, {Model Atmosphere Codes: ATLAS12 and ATLAS9}, ed.
  E.~{Niemczura}, B.~{Smalley}, \& W.~{Pych}, 39--51

\bibitem[{{Lang}(1993)}]{lang93}
{Lang}, K.~R. 1993, Astronomy, 21, 95

\bibitem[{{Lapenna} {et~al.}(2016){Lapenna}, {Lardo}, {Mucciarelli}, {Salaris},
  {Ferraro}, {Lanzoni}, {Massari}, {Stetson}, {Cassisi}, \&
  {Savino}}]{lapenna16}
{Lapenna}, E., {Lardo}, C., {Mucciarelli}, A., {et~al.} 2016, \apjl, 826, L1

\bibitem[{{Lind} {et~al.}(2012){Lind}, {Bergemann}, \& {Asplund}}]{lind12}
{Lind}, K., {Bergemann}, M., \& {Asplund}, M. 2012, \mnras, 427, 50

\bibitem[{{Luck}(2015)}]{luck15}
{Luck}, R.~E. 2015, \aj, 150, 88

\bibitem[{{Marino} {et~al.}(2011){Marino}, {Milone}, {Piotto}, {Villanova},
  {Gratton}, {D'Antona}, {Anderson}, {Bedin}, {Bellini}, {Cassisi}, {Geisler},
  {Renzini}, \& {Zoccali}}]{marino11}
{Marino}, A.~F., {Milone}, A.~P., {Piotto}, G., {et~al.} 2011, \apj, 731, 64

\bibitem[{{Marino} {et~al.}(2014){Marino}, {Milone}, {Przybilla}, {Bergemann},
  {Lind}, {Asplund}, {Cassisi}, {Catelan}, {Casagrande}, {Valcarce}, {Bedin},
  {Cort{\'e}s}, {D'Antona}, {Jerjen}, {Piotto}, {Schlesinger}, {Zoccali}, \&
  {Angeloni}}]{marino14}
{Marino}, A.~F., {Milone}, A.~P., {Przybilla}, N., {et~al.} 2014, \mnras, 437,
  1609

\bibitem[{{Merle} {et~al.}(2011){Merle}, {Th{\'e}venin}, {Pichon}, \&
  {Bigot}}]{merle11}
{Merle}, T., {Th{\'e}venin}, F., {Pichon}, B., \& {Bigot}, L. 2011, \mnras,
  418, 863

\bibitem[{{Mermilliod} {et~al.}(1997){Mermilliod}, {Mermilliod}, \&
  {Hauck}}]{merm97}
{Mermilliod}, J.-C., {Mermilliod}, M., \& {Hauck}, B. 1997, \aaps, 124, 349

\bibitem[{{Milone} {et~al.}(2013){Milone}, {Marino}, {Piotto}, {Bedin},
  {Anderson}, {Aparicio}, {Bellini}, {Cassisi}, {D'Antona}, {Grundahl},
  {Monelli}, \& {Yong}}]{milone13}
{Milone}, A.~P., {Marino}, A.~F., {Piotto}, G., {et~al.} 2013, \apj, 767, 120

\bibitem[{{Milone} {et~al.}(2012){Milone}, {Piotto}, {Bedin}, {King},
  {Anderson}, {Marino}, {Bellini}, {Gratton}, {Renzini}, {Stetson}, {Cassisi},
  {Aparicio}, {Bragaglia}, {Carretta}, {D'Antona}, {Di Criscienzo},
  {Lucatello}, {Monelli}, \& {Pietrinferni}}]{milone12}
{Milone}, A.~P., {Piotto}, G., {Bedin}, L.~R., {et~al.} 2012, \apj, 744, 58

\bibitem[{{Milone} {et~al.}(2010){Milone}, {Piotto}, {King}, {Bedin},
  {Anderson}, {Marino}, {Momany}, {Malavolta}, \& {Villanova}}]{milone10}
{Milone}, A.~P., {Piotto}, G., {King}, I.~R., {et~al.} 2010, \apj, 709, 1183

\bibitem[{{Milone} {et~al.}(2017){Milone}, {Piotto}, {Renzini}, {Marino},
  {Bedin}, {Vesperini}, {D'Antona}, {Nardiello}, {Anderson}, {King}, {Yong},
  {Bellini}, {Aparicio}, {Barbuy}, {Brown}, {Cassisi}, {Ortolani}, {Salaris},
  {Sarajedini}, \& {van der Marel}}]{milone17}
{Milone}, A.~P., {Piotto}, G., {Renzini}, A., {et~al.} 2017, \mnras, 464, 3636

\bibitem[{{Mucciarelli} {et~al.}(2017){Mucciarelli}, {Merle}, \&
  {Bellazzini}}]{muc17}
{Mucciarelli}, A., {Merle}, T., \& {Bellazzini}, M. 2017, \aap, 600, A104

\bibitem[{{Nardiello} {et~al.}(2015){Nardiello}, {Milone}, {Piotto}, {Marino},
  {Bellini}, \& {Cassisi}}]{nardiello15}
{Nardiello}, D., {Milone}, A.~P., {Piotto}, G., {et~al.} 2015, \aap, 573, A70

\bibitem[{{Piotto} {et~al.}(2007){Piotto}, {Bedin}, {Anderson}, {King},
  {Cassisi}, {Milone}, {Villanova}, {Pietrinferni}, \& {Renzini}}]{piotto07}
{Piotto}, G., {Bedin}, L.~R., {Anderson}, J., {et~al.} 2007, \apjl, 661, L53

\bibitem[{{Piotto} {et~al.}(2015){Piotto}, {Milone}, {Bedin}, {Anderson},
  {King}, {Marino}, {Nardiello}, {Aparicio}, {Barbuy}, {Bellini}, {Brown},
  {Cassisi}, {Cool}, {Cunial}, {Dalessandro}, {D'Antona}, {Ferraro}, {Hidalgo},
  {Lanzoni}, {Monelli}, {Ortolani}, {Renzini}, {Salaris}, {Sarajedini}, {van
  der Marel}, {Vesperini}, \& {Zoccali}}]{piotto15}
{Piotto}, G., {Milone}, A.~P., {Bedin}, L.~R., {et~al.} 2015, \aj, 149, 91

\bibitem[{{Prantzos} {et~al.}(2007){Prantzos}, {Charbonnel}, \&
  {Iliadis}}]{pr07}
{Prantzos}, N., {Charbonnel}, C., \& {Iliadis}, C. 2007, \aap, 470, 179

\bibitem[{{Prantzos} {et~al.}(2017){Prantzos}, {Charbonnel}, \&
  {Iliadis}}]{pr17}
{Prantzos}, N., {Charbonnel}, C., \& {Iliadis}, C. 2017, \aap, 608, A28

\bibitem[{{Prugniel} \& {Soubiran}(2001)}]{prusou01}
{Prugniel}, P. \& {Soubiran}, C. 2001, \aap, 369, 1048

\bibitem[{{Ruchti} {et~al.}(2013){Ruchti}, {Bergemann}, {Serenelli},
  {Casagrande}, \& {Lind}}]{ruchti13}
{Ruchti}, G.~R., {Bergemann}, M., {Serenelli}, A., {Casagrande}, L., \& {Lind},
  K. 2013, \mnras, 429, 126

\bibitem[{{Sbordone} {et~al.}(2011){Sbordone}, {Salaris}, {Weiss}, \&
  {Cassisi}}]{sbo11}
{Sbordone}, L., {Salaris}, M., {Weiss}, A., \& {Cassisi}, S. 2011, \aap, 534,
  A9

\bibitem[{{Seaton}(1979)}]{seaton79}
{Seaton}, M.~J. 1979, \mnras, 187, 73P

\bibitem[{{Sneden} {et~al.}(1992){Sneden}, {Kraft}, {Prosser}, \&
  {Langer}}]{sneden92}
{Sneden}, C., {Kraft}, R.~P., {Prosser}, C.~F., \& {Langer}, G.~E. 1992, \aj,
  104, 2121

\bibitem[{{Soto} {et~al.}(2017){Soto}, {Bellini}, {Anderson}, {Piotto},
  {Bedin}, {van der Marel}, {Milone}, {Brown}, {Cool}, {King}, {Sarajedini},
  {Granata}, {Cassisi}, {Aparicio}, {Hidalgo}, {Ortolani}, \&
  {Nardiello}}]{soto17}
{Soto}, M., {Bellini}, A., {Anderson}, J., {et~al.} 2017, \aj, 153, 19

\bibitem[{{Valdes} {et~al.}(2004){Valdes}, {Gupta}, {Rose}, {Singh}, \&
  {Bell}}]{valdes04}
{Valdes}, F., {Gupta}, R., {Rose}, J.~A., {Singh}, H.~P., \& {Bell}, D.~J.
  2004, \apjs, 152, 251

\bibitem[{{Ventura} {et~al.}(2001){Ventura}, {D'Antona}, {Mazzitelli}, \&
  {Gratton}}]{ventura01}
{Ventura}, P., {D'Antona}, F., {Mazzitelli}, I., \& {Gratton}, R. 2001, \apjl,
  550, L65

\bibitem[{{Ventura} {et~al.}(2013){Ventura}, {Di Criscienzo}, {Carini}, \&
  {D'Antona}}]{ventura13}
{Ventura}, P., {Di Criscienzo}, M., {Carini}, R., \& {D'Antona}, F. 2013,
  \mnras, 431, 3642

\bibitem[{{Villanova} {et~al.}(2009){Villanova}, {Piotto}, \&
  {Gratton}}]{villa09}
{Villanova}, S., {Piotto}, G., \& {Gratton}, R.~G. 2009, \aap, 499, 755

\bibitem[{{Yong} {et~al.}(2006){Yong}, {Aoki}, \& {Lambert}}]{yong06}
{Yong}, D., {Aoki}, W., \& {Lambert}, D.~L. 2006, \apj, 638, 1018

\bibitem[{{Yong} {et~al.}(2008){Yong}, {Grundahl}, {Johnson}, \&
  {Asplund}}]{yong08}
{Yong}, D., {Grundahl}, F., {Johnson}, J.~A., \& {Asplund}, M. 2008, \apj, 684,
  1159

\bibitem[{{Yong} {et~al.}(2005){Yong}, {Grundahl}, {Nissen}, {Jensen}, \&
  {Lambert}}]{yong05}
{Yong}, D., {Grundahl}, F., {Nissen}, P.~E., {Jensen}, H.~R., \& {Lambert},
  D.~L. 2005, \aap, 438, 875

\bibitem[{{Yong} {et~al.}(2015){Yong}, {Grundahl}, \& {Norris}}]{yong15}
{Yong}, D., {Grundahl}, F., \& {Norris}, J.~E. 2015, \mnras, 446, 3319

\end{thebibliography}
%%#####################################################################
%%#####################################################################

\newpage

%%%%%%%%%%%%%%%%%%%%%%%%%%%%%%%%%%%%%%%%%%%%%%%%%%%%%%%%%%%%%%%%%%%%%%%%%%%%%%%%%%%%%%%%%%%%%%%%%%%%%%%%%%%%%%%%%%%%%%%%%%%%%%%
\begin{appendix}

\end{appendix}

\end{document}